\documentclass[aps,prc,twocolumn,preprintnumbers,amsmath,amssymb,superscriptaddress,nofootinbib]{revtex4-1} 

\usepackage{amsthm}
\usepackage{mathtools}
\usepackage{physics}
\usepackage{graphicx}
\usepackage[left=23mm,right=13mm,top=35mm,columnsep=15pt]{geometry} 
\usepackage{adjustbox}
\usepackage{placeins}
\usepackage[T1]{fontenc}

\usepackage{graphicx}
\usepackage{dcolumn}
\usepackage{bm}
\usepackage{color}
\usepackage{threeparttable}
\usepackage{comment}
\usepackage{physics}
\usepackage[dvipsnames]{xcolor}
\usepackage[pdftex,colorlinks,citecolor=blue,urlcolor=blue,linkcolor=blue,bookmarks]{hyperref}
\begin{document}
\title{Correlations Between Neutrinoless Double-Beta, Double Gamow-Teller and Double-Magnetic Decays in the pnQRPA Framework}

\author{Lotta Jokiniemi}
    \email{ljokiniemi@triumf.ca}
\affiliation{TRIUMF, 4004 Wesbrook Mall, Vancouver, BC V6T 2A3, Canada}
\author{Javier Men\'{e}ndez}
    \email{menendez@fqa.ub.edu}    
\affiliation{Departament de Física Quàntica i Astrofísica, Universitat de Barcelona, 08028 Barcelona, Spain}
\affiliation{Institut de Ciències del Cosmos, Universitat de Barcelona, 08028 Barcelona, Spain}
\date{\today}

\begin{abstract}
We explore the relation between the nuclear matrix elements of neutrinoless double-beta ($0\nu\beta\beta$) decay and two other processes: double Gamow-Teller (DGT) and double-magnetic dipole (M1M1) transitions, with focus on medium-mass to heavy nuclei studied with the proton-neutron quasiparticle random-phase approximation (pnQRPA) framework. We explore a wide span of isoscalar proton-neutron pairing strengths
covering the typical range of values that describe well $\beta$- and two-neutrino $\beta\beta$-decay data.
Our results indicate good linear correlations between $0\nu\beta\beta$ and both DGT and M1M1 matrix elements. Together with future measurements of DGT and M1M1 transitions, these correlations could help constrain the values of the $0\nu\beta\beta$-decay nuclear matrix elements.
\end{abstract}

\maketitle

\section{Introduction}
\label{sec:introduction}
Observing neutrinoless double-beta ($0\nu\beta\beta$) decay, a hypothetical nuclear weak decay in which two neutrons in an atomic nucleus transform into two protons and only two electrons are emitted---we assume the energetically favored $\beta^-\beta^-$ mode---would be a clear signal of physics beyond the Standard Model of particle physics \cite{Avignone2008,Agostini2022}. Unlike any standard-model process, $0\nu\beta\beta$ decay changes
the number of leptons and of matter minus antimatter particles by two units. It also necessitates neutrinos to be \emph{Majorana particles}, in other words their own antiparticles. The potential to answer these fundamental physics questions drives extensive searches of $0\nu\beta\beta$ decay worldwide \cite{Majorana2022,Abe2023,Augier2022,Adams2021,Agostini2020,Anton2019,Azzolini2019,Abgrall2021,Adhikari2022,Adams2021b,Albanese2021,DARWIN:2020jme}.

The half-life of $0\nu\beta\beta$ decay ---the observable the experiments are after--- depends quadratically on nuclear matrix elements (NMEs) which are currently poorly known \cite{Agostini2022}. Furthermore, the hadronic two-body currents needed to reproduce experimental $\beta$-decay rates \cite{Pastore:2017uwc,Gysbers2019,King:2020wmp} can suppress $0\nu\beta\beta$-decay NMEs too \cite{Menendez2011,Jokiniemi2022}. On the other hand, Refs.~\cite{Cirigliano2018,Cirigliano2019} introduce a new short-range term to the $0\nu\beta\beta$-decay NMEs, which leads to a significant enhancement of the $^{48}$Ca NME~\cite{Wirth:2021pij}, and the impact in heavier $\beta\beta$ emitters may be similar~\cite{Jokiniemi2021}.

Nuclear structure measurements, such as two-nucleon processes, can be a valuable tool to shed light on the values of $0\nu\beta\beta$-decay NMEs \cite{Agostini2022}. Good examples are nucleon-pair transfer reactions~\cite{Brown14,Rebeiro20}
and two-neutrino double-beta ($2\nu\beta\beta$) decay, the standard-model-allowed $\beta\beta$ decay in which two antineutrinos are emitted alongside the electrons. The latter process has readily been measured in about a dozen nuclei \cite{Barabash2020}, and experimental half-lives are typically used to fix the proton-neutron pairing in many-body calculations using the proton-neutron quasiparticle random-phase approximation (pnQRPA) framework~\cite{Rodin2003,Rodin2005,Faessler2008,Simkovic2013,Simkovic2018,Mustonen:2013zu}.
Very recently, good correlations have been found between $2\nu\beta\beta$- and $0\nu\beta\beta$-decay NMEs~\cite{Horoi2022,Jokiniemi2022,Horoi2023}. These correlations allowed the prediction of $0\nu\beta\beta$-decay NMEs with theoretical uncertainties based on systematic nuclear shell-model (NSM) and pnQRPA calculations and measured $2\nu\beta\beta$-decay half-lives.

Similar studies have also been performed for other two-nucleon processes not yet measured. Double-charge-exchange reactions have raised a lot of interest as probes of $0\nu\beta\beta$-decay~\cite{NUMEN:2022ton}. These reactions open the possibility to access double Gamow-Teller (DGT) resonances and transitions to individual states~\cite{Takaki:2015pea,Takahisa:2017xry,Ejiri:2022ujl}. In fact, a very good linear correlation has been observed in the NSM between $0\nu\beta\beta$-decay and DGT NMEs to the ground state of the final nucleus~\cite{Shimizu2018}. The same correlation also holds for NMEs obtained with energy-density functional (EDF) theory~\cite{Rodriguez2013}. Moreover, calculations combining the NSM with variational Monte Carlo methods to capture additional short-range correlations via the generalized contact formalism~\cite{Weiss:2021rig}, and NMEs obtained with projected Hartree-Fock-Bogoliubov theory~\cite{Nautiyal:2022uew} also agree with the same $0\nu\beta\beta$-DGT correlation. \emph{Ab initio} studies using variational Monte Carlo, the no-core shell model, the in-medium generator coordinate method, and the valence-space in-medium similarity renormalization group (VS-IMSRG) approaches are also consistent with the linear correlation, which they find to be somewhat weaker~\cite{Yao22,Weiss:2021rig,Belley2022}. NMEs calculated with the interacting boson model (IBM-2) are also correlated~\cite{Santopinto2018}, but with a slope twice the one originally found for the NSM \cite{Barea2015,Brase2022}. In contrast, no correlation has been observed in the pnQRPA framework so far~\cite{Simkovic2011,Lv2023}. The absence of correlation has been related to the radial distribution of the DGT matrix elements~\cite{Shimizu2018} and to the spin-isospin SU(4) symmetry~\cite{Simkovic2018}. Very recently, in Ref. \cite{Lv2023}, the lack of correlation was attributed to the markedly different dependencies on the particle-hole and isovector particle-particle interactions of the DGT and $0\nu\beta\beta$-decay NMEs.

Another possibility to shed light on $0\nu\beta\beta$-decay NMEs is to study second-order electromagnetic transitions, which are connected to $\beta\beta$-decays through isospin symmetry~\cite{Ejiri68,Ejiri13}. Recently, the relation between $0\nu\beta\beta$- and double-gamma ($\gamma\gamma$) decays has been studied in the NSM framework~\cite{Romeo2021}. In order to favor the comparison, the authors calculated double magnetic-dipole (M1M1) decays in final $\beta\beta$ nuclei, in particular from an initial state which is the double isobaric analog of the initial $\beta\beta$ nuclei to the final ground state of the nucleus. Reference~\cite{Romeo2021} found a good linear correlation between $0\nu\beta\beta$-decay and $\gamma\gamma$-decay NMEs, with some dependence on the mass of the nuclei involved.

In this work, we investigate correlations between $0\nu\beta\beta$ decay, DGT and M1M1 transitions using the spherical pnQRPA framework performing systematic calculations with different proton-neutron pairing strength like in Ref.~\cite{Jokiniemi2022}. The study comprises nine $\beta\beta$-decay triplets, corresponding to the decays of $^{76}$Ge, $^{82}$Se, $^{96}$Zr, $^{100}$Mo, $^{116}$Cd, $^{124}$Sn, $^{128}$Te, $^{130}$Te and $^{136}$Xe.
We include the effect of two-body weak currents \cite{Menendez2011} and the short-range contribution \cite{Cirigliano2018,Cirigliano2019} into our $0\nu\beta\beta$-decay NMEs.

\section{Nuclear transitions}
\label{sec:formalism}

\subsection{Neutrinoless Double-Beta Decay}
Assuming that light Majorana-neutrino exchange is the dominant $0\nu\beta\beta$-decay mechanism, we can write the half-life of the
decay as
\begin{equation}
[t_{1/2}^{0\nu}]^{-1}=G_{0\nu}\,g_{\rm A}^4\,|M^{0\nu}_{\rm L}+M^{0\nu}_{\rm S}|^2\left(\frac{m_{\beta\beta}}{m_e}\right)^2\;,
  \label{eq:half-life}
\end{equation}
where $G_{0\nu}$ is a phase-space factor for the final-state leptons \cite{Kotila2012}, $M_{\rm L}^{0\nu}$ and $M_{\rm S}^{0\nu}$ are the long- and short-range components of the $0\nu\beta\beta$ NME, respectively, and $g_A=1.27$ is the axial coupling constant. The term $m_{\beta\beta}=\sum_{j={\rm light}}(U_{ej})^2m_j$ characterises the lepton-number violation through the two additional physical Majorana phases. $U$ is the neutrino mixing matrix and $m_j$, $m_e$ the neutrino and electron masses, respectively.
The standard long-range NME can be decomposed as
\begin{equation}
    M^{0\nu}_{\rm L}=M_{\rm GT}^{0\nu}-M_{F}^{0\nu}+M_{\rm T}^{0\nu}\;,
    \label{eq:0vbb-NME}
\end{equation}
where $M_{\rm GT}^{0\nu}$, $M_{\rm F}^{0\nu}$ and $M_{\rm T}^{0\nu}$ are the Gamow-Teller, Fermi and tensor parts.

In the pnQRPA framework, the matrix elements $M_K^{0\nu}$, $K={\rm F, GT, T}$, are computed without resorting to the closure approximation:
\begin{equation}
\begin{split}
M_K^{0\nu}=&\sum_{J^{\pi}k_1k_2\mathcal{J}}\sum_{pp'nn'}(-1)^{j_n+j_{p'}+J+\mathcal{J}}\widehat{\mathcal{J}}\begin{Bmatrix}
    j_p & j_n & J\\
    j_{n'} &j_{p'} &\mathcal{J}
    \end{Bmatrix}\\
&\times(pp':\mathcal{J}||H_{K}(r,E_k)f^2_{\rm SRC}(r)\mathcal{O}_K\tau^-_1\tau^-_2||nn':\mathcal{J})\\
    &\times (0_f^+||[c_{p'}^{\dag}\tilde{c}_{n'}]_J||J_{k_1}^{\pi})\langle J_{k_1}^{\pi}|J_{k_2}^{\pi}\rangle(J_{k_2}^{\pi}||[c_{p}^{\dag}\tilde{c}_{n}]_J||0_i^+)\;,
    \end{split}
    \label{eq:pnQRPA-NMEs}
\end{equation}
where $\widehat{\mathcal{J}}=\sqrt{2\mathcal{J}+1}$ and $r=|\mathbf{r}_1-\mathbf{r}_2|$ is the distance between the neutrons $(n,n')$ which decay into protons $(p,p')$. Nucleons populate single-particle orbitals with total angular momentum $j_n,j_p$, and the isospin operator $\tau^-$ brings neutrons into protons. $k_1$ ($k_2$) labels the different pnQRPA solutions for a given total angular-momentum-parity $J^{\pi}$ based on the final (initial) nucleus of the decay, and $E_k$ is their average energy.

For the long-range NME, the operators $\mathcal{O}_K$ are the
Fermi, Gamow-Teller and tensor operators
    $\mathcal{O}_{\rm F}=\boldsymbol{1}$,
    $\mathcal{O}_{\rm GT}=\boldsymbol{\sigma}_1\cdot\boldsymbol{\sigma}_2$, and 
    $\mathcal{O}_{\rm T}
    =3[(\boldsymbol{\sigma}_1\cdot\hat{\mathbf{r}}_{12})(\boldsymbol{\sigma}_2\cdot\hat{\mathbf{r}}_{12})]-\boldsymbol{\sigma}_1\cdot\boldsymbol{\sigma}_2$.
The neutrino potential $H_{K}$ is defined as
\begin{equation}
    H_K(r,E_k)=\frac{2R}{\pi g_{\rm A}^2}\int_0^{\infty}\frac{p\,h_K(p^2)j_{\lambda}(pr){\rm d}p}{p+E_k-(E_i+E_f)/2}\;,
    \label{eq:H_K}
\end{equation}
where $E_i,E_f$ denote the energy of the initial and final nuclei, the function $f_{\rm SRC}$ takes into account short-range correlations and $j_{\lambda}$ is the spherical Bessel function with $\lambda=0$ for $K=\text{F, GT}$ and $\lambda=2$ for $K=\text{T}$. The radius $R=1.2A^{1/3}$, where $A$ is the nuclear mass number, is introduced to make the NMEs dimensionless. The $h$ functions contain the weak couplings and depend on the momentum transfer $p$. The dominant GT term reads
\begin{align}
    h_{\rm GT}=&g^2_{\rm A}(p^2)-\frac{g_{\rm A}(p^2)g_{\rm P}(p^2)p^2}{3m_{\rm N}}+\frac{g^2_{\rm P}(p^2)p^4}{12m_{\rm N}^2}
    +\frac{g^2_{\rm M}(p^2)p^2}{6m_{\rm N}^2}\;.
    \label{eq:h_gt2}
\end{align}
The axial and magnetic couplings $g_{\rm A}(p^2)$ and $g_{\rm M}(p^2)$ include the usual dipole form factor with axial-vector~\cite{Bernard2001} and vector \cite{Dumbrajs1983} masses, correspondingly. The pseudoscalar coupling is $g_{\rm P}(p^2)=2m_{\rm N}g_{\rm A}(p^2)(p^2+m_{\pi}^2)^{-1}$, with $m_{\rm N}$ and $m_{\pi}$ the nucleon and pion masses. The other $h$ terms have similar forms and are given in Ref.~\cite{Engel2017}.

For the short-range NME, in the pnQRPA we sum over intermediate states in Eq. \eqref{eq:pnQRPA-NMEs} with the operator $\mathcal{O}_{\rm S}=\mathcal{O}_{\rm F}$ and the neutrino potential
\begin{equation}
    H_S(r)=\frac{2R}{\pi g_{\rm A}^2}\int j_0(pr)h_{\rm S}(p^2)p^2{\rm d}p\;,
\end{equation}
where $h_{\rm S}(p^2)=2g_{\nu}^{\rm NN}e^{-p^2/(2\Lambda ^2)}$ with Gaussian regulator scale $\Lambda$ and coupling $g_{\nu}^{\rm NN}$ taken from the charge-independence-breaking terms of different Hamiltonians as in Ref.~\cite{Jokiniemi2021}. This assumes that the two relevant couplings entering charge-independence breaking are equal, a relatively good approximation supported by quantum chromodynamics calculations using dispersion relations~\cite{Cirigliano:2020dmx,Cirigliano:2021qko} and large number of colors~\cite{Richardson:2021xiu}. More definitive $g_{\nu}^{\rm NN}$ determinations obtained with lattice quantum chromodynamics techniques are in progress~\cite{Davoudi:2020gxs,Davoudi:2021noh}.

In addition, we approximate chiral-effective-field-theory two-body weak currents (2BCs) as effective one-body operators via normal ordering with respect to a spin-isospin symmetric Fermi gas reference state~\cite{Menendez2011,Engel:2014pha,Klos2013,Hoferichter2020}. This leads to the replacement 
\begin{align}
  g_{\rm A}(p^2,{\rm 2b})&\rightarrow g_{\rm A}(p^2)+\delta_a(p^2)\,,  \\
  g_{\rm P}(p^2,{\rm 2b})&\rightarrow g_{\rm P}(p^2)-\frac{2m_{\rm N}}{p^2}\,\delta_a^P(p^2)\,,
\end{align}
with two-body functions $\delta_a(p^2)$, $\delta_a^P(p^2)$ dependent on the Fermi-gas density $\rho$ and chiral-effective-field-theory low-energy couplings. We take the same values for these as in Ref.~\cite{Hoferichter2020}. As for $\beta$ decay, normal-ordered currents approximate well the full two-body results~\cite{Gysbers2019}.

The one-body transition densities between the initial (final) $0^+$ ground state and a given $J^{\pi}_k$ state in the intermediate odd-odd nucleus are obtained from
\begin{equation}
\begin{split}
    (J_{k_2}^{\pi}||[c_{p}^{\dag}\tilde{c}_{n}]_J||0_i^+)&=\widehat{J}[u_pv_nX_{pn}^{J^{\pi}k_2}+v_pu_nY_{pn}^{J^{\pi}k_2}]\;,\\
    (0_f^+||[c_{p'}^{\dag}\tilde{c}_{n'}]_J||J_{k_1}^{\pi})&=\widehat{J}[\bar{v}_{p'}\bar{u}_{n'}\bar{X}_{p'n'}^{J^{\pi}k_1}+\bar{u}_{p'}\bar{v}_{n'}Y_{p'n'}^{J^{\pi}k_1}]\;,
\end{split}
    \label{eq:pnQRPA-OBTDs}
\end{equation}
where $v(\bar{v})$ and $u(\bar{u})$ are the BCS occupation and vacancy amplitudes of the initial (final) even-even nucleus. The $X(\bar{X})$ and $Y(\bar{Y})$ are the forward- and backward-going amplitudes emerging from the pnQRPA calculation based on the initial (final) nucleus. The overlap between the two sets of $J^{\pi}$ states $\langle J_{k_1}^{\pi}|J_{k_2}^{\pi}\rangle$ can be written as
\begin{equation}
    \langle J_{k_1}^{\pi}|J_{k_2}^{\pi}\rangle=\sum_{pn}\left[X_{pn}^{J_{k_1}^{\pi}}\bar{X}_{pn}^{J_{k_2}^{\pi}}-Y_{pn}^{J_{k_1}^{\pi}}\bar{Y}_{pn}^{J_{k_2}^{\pi}}\right]\;.
    \label{eq:overlap}
\end{equation}

\subsection{Two-Neutrino Double-Beta Decay}

The $2\nu\beta\beta$-decay half-life can be written in the form
\begin{equation}
 [t_{1/2}^{2\nu}]^{-1}=g_{\rm A}^4G_{2\nu}\,|M^{2\nu}|^2\;,
  \label{eq:2vbb-half-life}   
\end{equation}
where $G_{2\nu}$ is the phase-space factor~\cite{Kotila2012}.  Note that instead of $g_A$, an effective coupling $g_{\rm A}^{\rm eff}=q\,g_{\rm A}$ is often introduced, where $q$ is a quenching factor needed to reproduce measured $2\nu\beta\beta$ rates. Here we use the bare $g_A$ value with $q=1$ unless otherwise specified.
$M^{2\nu}$ is the $2\nu\beta\beta$-decay NME:
\begin{equation}
M^{2\nu}=M_{\rm GT}^{2\nu}+M_{\rm F}^{2\nu}\;,
\label{eq:m_2v}
\end{equation}
with Gamow-Teller and Fermi parts. Because isospin is a good quantum number in nuclei, the $2\nu\beta\beta$-decay Fermi matrix element should approximately vanish. Thus, in the pnQRPA calculations we force $M_{\rm F}^{2\nu}\approx 0$ in order to restore isospin symmetry \cite{Simkovic2013}. The remaining NME is calculated as
\begin{equation}
\begin{split}
    M^{2\nu}_{\rm GT}=&\sum_{k_1,k_2}(0^+_{ f}||\sum_a\tau^-_a\boldsymbol{\sigma}_a||1^+_{k_1})\langle 1^+_{k_1}|1^+_{k_2} \rangle\\
    &\times(1^+_{k_2} ||\sum_b\tau^-_b\boldsymbol{\sigma}_b||0^+_{ i})/D_{k}\;,
    \end{split}
    \label{eq:M^2v_GT}
\end{equation}
where $D_{k}=(E_k-(E_i+E_f)/2)/m_e$. The overlap $\langle 1^+_{k_1}|1^+_{k_2} \rangle$ is defined in Eq. \eqref{eq:overlap}.
In the pnQRPA, one-body matrix elements for an operator $\mathbf{O}$ with rank $L$, like those in Eq. \eqref{eq:M^2v_GT} where $\mathbf{O}_L=\boldsymbol{\sigma}$, can be obtained from the general formulae
\begin{align}
  (J_{k_2}^{\pi}||\sum_a \tau^-_a\mathbf{O}_{L,a}||0^+_i)=&\delta_{LJ}\frac{1}{\widehat{L}}\sum_{pn}(p||\mathbf{O}_L||n)\nonumber\\
   &\times(J_{k_2}^{\pi}||[c_{p}^{\dag}\tilde{c}_{n}]_J||0_i^+)\label{eq:one-body_NME}\;,\\
   (0^+_f||\sum_a \tau^-_a\mathbf{O}_{L,a}||J^{\pi}_{k_1})=&\delta_{LJ}\frac{1}{\widehat{L}}\sum_{p'n'}(p'||\mathbf{O}_L||n')\nonumber\\
   &\times(0_f^+||[c_{p'}^{\dag}\tilde{c}_{n'}]_J||J_{k_1}^{\pi})\;.
   \label{eq:one-body_NME2}
   \end{align}

For completeness, the (vanishing) Fermi part of the $2\nu\beta\beta$-decay NME is given by
\begin{equation}
\begin{split}
    M^{2\nu}_{\rm F}=&\Big(\frac{g_{\rm V}}{g_{\rm A}}\Big)^2\sum_{k_1,k_2}(0^+_{f}||\sum_a\tau^-_a||0^+_{k_1})\langle 0^+_{k_1}|0^+_{k_2} \rangle\\
    &\times(0^+_{k_2} ||\sum_b\tau^-_b||0^+_{i})/D_k\;,
    \end{split}
\end{equation} 
with the vector coupling $g_{\rm V}=1.0$ and $D_k$ depending now on the energies of the intermediate $0^+$ states. The one-body NMEs can again be obtained from Eqs.~\eqref{eq:one-body_NME} and~\eqref{eq:one-body_NME2}  with $\mathbf{O}_L=\mathbf{1}$.

\subsection{Beta Decay and Electron Capture}
The $\log ft$ value for a $\beta$ decay or an electron capture (EC) can be written as \cite{Suhonen2007}
\begin{equation}
    \log ft=\log (f_0t_{1/2}[{\rm s}])=\log\left(\frac{\kappa}{B_{\rm F}+B_{\rm GT}}\right)\;,
    \label{eq:logft}
\end{equation}
where $\kappa=2\pi^2\hbar^7\ln 2/(m_e^2G_F^2)\approx 6289~{\rm s}$ \cite{ParticleDataGroup:2016lqr} and $B_{\rm F}$ and $B_{\rm GT}$ are Fermi and Gamow-Teller reduced transition probabilities:
\begin{align}
    B_{\rm F}=&\frac{g_{\rm V}^2}{2J_i+1}|M_{\rm F}|^2\;,\\
    B_{\rm GT}=&\frac{g_{\rm A}^2}{2J_i+1}|M_{\rm GT}|^2\;,
\end{align}
where $J_i$ is the angular momentum of the initial state and the nuclear matrix elements are defined as
\begin{align}
    M_{\rm F}=&(J^{\pi_f}_{f}||\sum_a\tau^-_a||J^{\pi_i}_{i})\;,\\
    M_{\rm GT}=&(J^{\pi_f}_{f}||\sum_a\tau^-_a\boldsymbol{\sigma}_a||J^{\pi_i}_{i})\;,
\end{align}
where $\pi_i=\pi_f$, since the allowed F and GT transitions are parity-conserving.
All transitions considered in the present work involve a nucleus $0^+$ ground state. In that case, the required one-body matrix elements are given by Eqs. \eqref{eq:one-body_NME} and \eqref{eq:one-body_NME2}.

\subsection{Double Gamow-Teller Transitions}

We define the DGT NME as 
\begin{equation}
    M_{\rm DGT}=-\langle 0^+_{f}||\sum_{a,b}[\boldsymbol{\sigma}_a\tau^-_a\otimes\boldsymbol{\sigma}_b\tau^-_b]^0||0^+_{i}\rangle\;,
\end{equation}
that is proportional to the $2\nu\beta\beta$-decay NME between the same states calculated in the closure approximation \cite{Varshalovich1988}:
\begin{align}
M_{\rm DGT}&=\frac{1}{\sqrt{3}}M^{2\nu}_{\rm GTcl}\;,
    \label{eq:DGT_2v-closure_relation} \\
    M^{2\nu}_{\rm GTcl}&=\langle 0^+_{f}|\sum_{a,b}\tau^-_a\tau^-_b\boldsymbol{\sigma}_a\cdot\boldsymbol{\sigma}_b|0^+_{i}\rangle
    \label{eq:M_GTcl}\,. 
\end{align}
Note that, in contrast with the definition in Ref.~\cite{Shimizu2018}, here we can have positive or negative $M_{\rm DGT}$ values.

In the pnQRPA framework, the closure $2\nu\beta\beta$-decay NME is obtained as \cite{Simkovic2018}
\begin{equation}
\begin{split}
    M^{2\nu}_{\rm GTcl}=&\sum_{k_1,k_2}(0^+_{f}||\sum_a\tau^-_a\boldsymbol{\sigma}_a||1^+_{k_1})\langle 1^+_{k_1}|1^+_{k_2} \rangle\\
    &\times(1^+_{k_2} ||\sum_b\tau^-_b\boldsymbol{\sigma}_b||0^+_{i})\;.
    \end{split}
    \label{eq:GTcl-pnQRPA}
\end{equation}
Combining Eqs. \eqref{eq:DGT_2v-closure_relation} and \eqref{eq:GTcl-pnQRPA} we arrive at the expression for the DGT matrix element:
\begin{equation}
 \begin{split}
    M_{\rm DGT}=&\frac{1}{\sqrt{3}}\sum_{k_1,k_2}(0^+_{f}||\sum_a\tau^-_a\boldsymbol{\sigma}_a||1^+_{k_1})\langle 1^+_{k_1}|1^+_{k_2} \rangle\\
    &\times(1^+_{k_2} ||\sum_b\tau^-_b\boldsymbol{\sigma}_b||0^+_{i})\;.
    \end{split}  
    \label{eq:DGT-pnQRPA}
\end{equation}

Likewise, the Fermi NME in the closure approximation is calculated as
\begin{equation}
\begin{split}
    M^{2\nu}_{\rm Fcl}=&\Big(\frac{g_{\rm V}}{g_{\rm A}}\Big)^2\sum_{k_1,k_2}(0^+_{f}||\sum_a\tau^-_a||0^+_{k_1})\langle 0^+_{k_1}|0^+_{k_2} \rangle\\
    &\times(0^+_{k_2} ||\sum_b\tau^-_b||0^+_{i})\;.
    \end{split}
\end{equation}

\subsection{Double-Magnetic Dipole Gamma Decay}

The magnetic dipole (M1) operator is defined as
\begin{equation}
    \mathbf{M1}=\mu_N\sqrt{\frac{3}{4\pi}}\sum_{a=1}^A( g_i^l\boldsymbol{\ell}_i+g_i^s\mathbf{s}_i)\;,
\end{equation}
where $\mu_N$ is the nuclear magneton, and the neutron and
proton spin and orbital $g$-factors are: $g_s^n=-3.826$, $g_s^p=5.586$, $g_l^n=0$, and $g_l^p=1$
(note that using effective $g$-factors does not affect much the correlation with $0\nu\beta\beta$-decay NMEs in NSM calculations~\cite{Romeo2021}).
Furthermore, $\mathbf{s}=\tfrac12\boldsymbol{\sigma}$ is the spin operator and $\boldsymbol{\ell}$ the orbital-angular-momentum operator. For the most probable case of two photons emitted with the same energy, the NME for double-magnetic dipole (M1M1) transition can be written as \cite{Romeo2021}
\begin{equation}
    M^{\gamma\gamma}({\rm M1M1})
    =\sum_k\frac{( 0^+_{f}||\mathbf{M1}||1^+_k)( 1^+_k||\mathbf{M1}||0^+_{i})}{E_k-(E_i+E_f)/2}\;.
    \label{eq:M_M1M1}
\end{equation}

We study the M1M1 $\gamma\gamma$-transition from the double-isobaric analog state (DIAS) of the ground state of a $\beta\beta$ emitter to the ground state of the final nucleus of a $\beta\beta$-decay triplet. This DGT transition can be related to the ground-state-to-ground-state $\beta\beta$-decay, since in both cases the isospin of the initial and final states are related by $T_i=T_f+2$. Moreover, for the same operator both processes share a common reduced matrix element in isospin space, with non-reduced matrix elements related by a factor $\alpha=\tfrac12\sqrt{(2+T_f)(3+2T_f)}$, given by the Wigner-Eckart theorem~\cite{Edmonds57}. Even though the operators are different, Ref.~\cite{Romeo2021} finds a good correlation between M1M1 transitions from the DIAS (multiplied by $\alpha$) and $0\nu\beta\beta$-decay NMEs.

DIAS are not well described in the QRPA formalism because it does not conserve isospin symmetry. Alternatively, we obtain $\alpha M^{\gamma\gamma}$ by its isospin-rotated equivalent, calculating the M1M1-decays as charge-changing transitions between the different isotopes (quite like $2\nu\beta\beta$ decay) in the pnQRPA formalism as
\begin{equation}
\begin{split}
    \alpha M^{\gamma\gamma}({\rm M1M1})=&\mu_{\rm N}^2 \frac{3}{4\pi}\sum_{k_1,k_2}\frac{\langle 1^+_{k_1}|1^+_{k_2} \rangle}{E_k-(E_i+E_f)/2}\\
    &\times(0^+_{f}||\sum_a\tau^-_a(g_l^{T=1}\boldsymbol{\ell}_a+g_s^{T=1}\mathbf{s}_a)||1^+_{k_1})\\
    &\times(1^+_{k_2} ||\sum_b\tau^-_b(g_l^{T=1}\boldsymbol{\ell}_b+g_s^{T=1}\mathbf{s}_b)||0^+_{i})\;,
    \end{split}
    \label{eq:M_M1M1_QRPA}
\end{equation}
where $g_l^{T=1}=\tfrac12(g_{\rm n}^l-g_{\rm p}^l)$ and $g_s^{T=1}=\tfrac12(g_{\rm n}^s-g_{\rm p}^s)$ are the isovector ($T=1$) angular-momentum and spin $g$-factors. The one-body NMEs can be obtained from Eqs. \eqref{eq:one-body_NME} and \eqref{eq:one-body_NME2} by substituting $\mathbf{O}_L=g_l^{T=1}\boldsymbol{\ell}+g_s^{T=1}\mathbf{s}$. Our calculation therefore neglects the possible mixing of different isospin components in the initial DIAS state, a common limitation with the NSM study~\cite{Romeo2021}.

\section{pnQRPA}

\begin{table}[t]
    \centering
    \caption{Values of $g_{\rm pp}^{T=1}$ adjusted so that the non-closure (second column) or closure (third column) Fermi $2\nu\beta\beta$-decay NME vanishes, and values of $g_{\rm pp}^{T=0}$ adjusted to reproduce the measured $2\nu\beta\beta$-decay half-life \cite{Barabash2020} with $1.0\leq g_{\rm A}^{\rm eff}\leq 1.27$ (fourth column) or so that the closure Gamow-Teller NME vanishes (fifth column). Values in brackets indicate that adjusting to the measured half-life is not possible (see text).}
    \begin{ruledtabular}
    \begin{tabular}{ccccc}
     &\multicolumn{2}{c}{$g_{\rm pp}^{T=1}$} &\multicolumn{2}{c}{$g_{\rm pp}^{T=0}$}\\
     \cline{2-3}\cline{4-5}
     Nucleus & $M_{\rm F}^{2\nu}=0$ & $M_{\rm Fcl}^{2\nu}=0$ & $M_{\rm GT}^{2\nu}=M_{\rm exp}^{2\nu}$ &$M_{\rm GTcl}^{2\nu}=0$\\
     \hline
        $^{76}$Ge &0.96 &0.96 &$0.83-0.85$ &0.83\\
        $^{82}$Se &0.95 &0.95 &$0.82-0.83$ &0.81\\
        $^{96}$Zr &0.92 &0.91 &(0.83) &0.83\\
        $^{100}$Mo &0.91 &0.91 &$0.87-0.89$ &0.80\\
        $^{116}$Cd &0.82 &0.82 &$0.82-0.85$ &0.87\\
        $^{124}$Sn &0.85 &0.83 &(0.75) &0.71\\
        $^{128}$Te &0.87 &0.87 &$0.75-0.76$ &0.72\\
        $^{130}$Te &0.86 &0.85 &$0.73-0.74$ &0.70\\
        $^{136}$Xe &0.86 &0.85 &$0.67-0.69$ &0.68\\
    \end{tabular}
    \end{ruledtabular}
    \label{tab:gpps-2vbb}
\end{table}

We use the spherical proton-neutron QRPA in large no-core single-particle bases comprising 18 orbitals for $A=76,82$ systems, 25 orbitals for $A=96,100$ systems, and 26 orbitals for $A=116,124,128,130$ and 136 systems. The same orbitals are used for both protons and neutrons. These bases span all the orbits from the $n=0$ oscillator major shell up to at least two oscillator major shells above the respective Fermi level for protons and neutrons. Single-particle energies are obtained by solving the radial Schr\"{o}dinger equation for a Coulomb-corrected Woods-Saxon potential optimized for nuclei close to $\beta$-stability~\cite{Bohr1969}. The resulting proton and neutron single-particle energies of the orbitals close to the Fermi surfaces have been slightly modified in order to better reproduce the low-lying spectra of the neighboring odd-mass nuclei. The single-particle bases correspond to those used in previous $0\nu\beta\beta$-decay and ordinary-muon-capture studies \cite{Jokiniemi2018,Jokiniemi2019}, apart from the $A=124$ system not included in these works.

\begin{table*}[t]
    \centering
    \caption{Measured $\log ft$ values for Gamow-Teller transitions involving nuclei in $\beta\beta$ triplets together with the corresponding NME ($g_{\rm A}^{\rm eff}M_{\rm GT}$) and adjusted $g_{\rm pp}^{T=0}$ values. The ranges come from variation of the axial-vector coupling $1.0\leq g_{\rm A}^{\rm eff}\leq 1.27$.}
    \begin{ruledtabular}
    \begin{tabular}{cccccc}
     $A$  &Decay& Type &$\log ft_{\rm exp}$ \cite{NNDC}  &$g_{\rm A}^{\rm eff}M_{\rm GT}$ &$g_{\rm pp}^{T=0}$\\
     \hline82 &${\rm Rb}(1^+_{\rm gs})\rightarrow {\rm Kr}(0^+_{\rm gs})$ &EC &4.576 &0.708 &$0.38-0.46$\\
     100 & ${\rm Tc}(1^+_1)\rightarrow {\rm Mo}(0^+_{\rm gs})$ &EC &4.3  &0.972 &$0.50-0.63$\\
     100 &${\rm Tc}(1^+_1)\rightarrow {\rm Ru}(0^+_{\rm gs})$ &$\beta^-$ &4.598 &0.690 &$0.88-0.89$\\
     128 &${\rm I}(1^+_1)\rightarrow {\rm Xe}(0^+_{\rm gs})$ &$\beta^-$ &6.061 &0.128 &$0.73-0.75$\\
    \end{tabular}
    \end{ruledtabular}
    \label{tab:gpps-b}
\end{table*}

The quasiparticle spectra, needed in the pnQRPA diagonalization, are obtained by solving the BCS equations for protons and neutrons, separately. We use the two-body interaction derived from the Bonn-A one-boson exchange potential \cite{Holinde1981} and fine-tune it by adjusting the proton and neutron pairing parameters to the phenomenological pairing gaps extracted from proton and neutron separation energies. The resulting pairing gaps and pairing strengths are tabulated in Refs.~\cite{Jokiniemi2018,Jokiniemi2019} except for the $A=124$ system where
\begin{equation}
\begin{cases}
g_{\rm pair}^{\rm (p)}=0.858 \;,g_{\rm pair}^{\rm (n)}=0.836 \text{ for } ^{124}{\rm Sn}\,,\\
g_{\rm pair}^{\rm (p)}=0.824 \;,g_{\rm pair}^{\rm (n)}=0.812 \text{ for } ^{124}{\rm Te}\,.
\end{cases}
\end{equation}

The residual Hamiltonian for the pnQRPA calculation contains two adjustable factors: the particle-hole parameter $g_{\rm ph}$ scaling the particle-hole channel, and the particle-particle parameter $g_{\rm pp}$ scaling the particle-particle channel. We fix $g_{\rm ph}$ to reproduce the centroid of the Gamow-Teller giant resonance in calculations for the $1^+$ channel. The particle-particle parameter is known to have a strong influence on nuclear operators driven by the nuclear spin \cite{Suhonen2005,Faessler2008,Jokiniemi2018} and requires a closer look.

Traditionally, $g_{\rm pp}$ has been adjusted to $2\nu\beta\beta$-decay data, whenever possible. This method has later been replaced by the so-called partial isospin-restoration scheme \cite{Simkovic2013}, where $g_{\rm pp}$ is divided into isoscalar ($T=0$) and isovector ($T=1$) parts, which scale the $T=0$ and $T=1$ particle-particle channels of the pnQRPA matrix, accordingly. The $T=1$ part, $g_{\rm pp}^{T=1}$, is adjusted so that the Fermi part of the $2\nu\beta\beta$-decay NME vanishes, restoring isospin symmetry. The remaining isoscalar part, $g_{\rm pp}^{T=0}$, is then independently fixed to reproduce the experimental $2\nu\beta\beta$-decay half-life for a given value of the effective axial-vector coupling $g_{\rm A}^{\rm eff}$.
More recently, Ref.~\cite{Simkovic2018} has proposed to restore the spin-isospin SU(4) symmetry by forcing $M^{2\nu}_{\rm GTcl}\approx0$ by adjusting the value of $g_{\rm pp}^{T=0}$, and to fix $g_{\rm pp}^{T=1}$ so that $M^{2\nu}_{\rm Fcl}\approx0$. 

Table \ref{tab:gpps-2vbb} presents the $g_{\rm pp}$ values we obtained from the different adjustment methods. For the cases where the $2\nu\beta\beta$-decay half-life has not been measured ($^{124}$Sn) or cannot be reproduced by our pnQRPA setup ($^{96}$Zr), we give a value at a safe distance ($\sim 0.01$) before the $2\nu\beta\beta$ NME becomes unstable \cite{Suhonen2005,Faessler2008,Suhonen2014}. Not in all cases the same value of $g_{\rm pp}^{T=1}$ gives both $M^{2\nu}_{\rm F}\approx0$ and $M^{2\nu}_{\rm Fcl}\approx0$ --- most notably for $^{124}$Sn the two values differ by 0.02. The differences are due to the energy denominator in the non-closure NME, which mitigates high-energy contributions. For $g_{\rm pp}^{T=0}$ the variation between the two different adjustment methods can be even more significant (e.g. for $^{100}$Mo and $^{130}$Te). This is again partly because the energy denominator in $M^{2\nu}_{\rm GT}$, but mainly due to the fact that restoring SU(4) does not reproduce exactly $2\nu\beta\beta$-decay half-lives~\cite{Simkovic2018}.

Alternatively we can adjust $g_{\rm pp}^{T=0}$ to measured single-$\beta$ decays or ECs~\cite{Aunola1996,Suhonen2005,Marketin2016,Hinohara:2022uip,Popara:2021lst}. Table \ref{tab:gpps-b} lists experimental $\log ft$ values for a set of $\beta^-$ and EC Gamow-Teller transitions together with the corresponding NMEs and adjusted $g_{\rm pp}^{T=0}$ values. We vary the effective value of the axial coupling $1.0\leq g_{\rm A}^{\rm eff}\leq 1.27$. Note that the $\beta^-$ and EC NMEs behave in opposite ways as function of $g_{\rm pp}$: while $\beta^-$-decay NMEs decrease, EC ones increase with higher values of $g_{\rm pp}^{T=0}$ \cite{Suhonen2005,Faessler2008}. Thus, reproducing  $\beta$-decay (EC) data requires large (small) $g_{\rm pp}^{T=0}$ values, see Table \ref{tab:gpps-b}.
 
It is noteworthy that in all the above-mentioned adjustment methods $g_{\rm pp}^{T=0}$ and $g_{\rm A}^{\rm eff}$ are strongly correlated. Ideally, the same choice of parameter set ($g_{\rm A}^{\rm eff}$, $g_{\rm pp}^{T=0}$) would reproduce all three observables. However, the values presented in Tables \ref{tab:gpps-2vbb} and \ref{tab:gpps-b} suggest that this is not the case; 
 the values of $g_{\rm pp}^{T=0}$ needed to reproduce $2\nu\beta\beta$-decay data normally fail to reproduce $\beta^-$/EC data and vice versa. This shortcoming of the pnQRPA framework has already been pointed out in previous analyses, which indicate that reproducing both EC and $\beta$-decay $\log ft$ values and $\beta\beta$-decay half-lives with a given $g_{\rm pp}$ parameter typically requires different effective $g_{\rm A}$ values for $\beta$ and $\beta\beta$ decays \cite{Suhonen2013,Suhonen2014}. Alternatively, to simultaneously reproduce all three observables often requires small $g_{\rm pp}$ and effective $g_{\rm A}$ values \cite{Faessler2008}. 
 
In order to take this uncertainty into account, in the present study we consider the conservative range $0.6\leq g_{\rm pp}^{T=0}\leq 0.8$ (except for $^{124}$Sn since $g_{\rm pp}^{T=0}=0.8$ is beyond the pnQRPA breakdown \cite{Suhonen2005,Suhonen2007,Faessler2008,Suhonen2014}) and use the bare value $g_{\rm A}=1.27$. For the isovector part $g_{\rm pp}^{T=1}$ we use the value that restores isospin symmetry: $M_{\rm F}^{2\nu}=0$ (column 2 in Table \ref{tab:gpps-2vbb}). In addition, we include the results obtained with $g_{\rm pp}$ adjusted via the partial isospin-restoration scheme
(columns 2 and 4 in Table \ref{tab:gpps-2vbb}), the most common way to adjust $g_{\rm pp}^{T=0}$ in recent $0\nu\beta\beta$-decay studies \cite{Simkovic2013,Hyvarinen2015}.

\section{Correlation between DGT and $0\nu\beta\beta$-decay NMEs}

\begin{figure}[t]
    \centering
    \includegraphics[width=\linewidth]{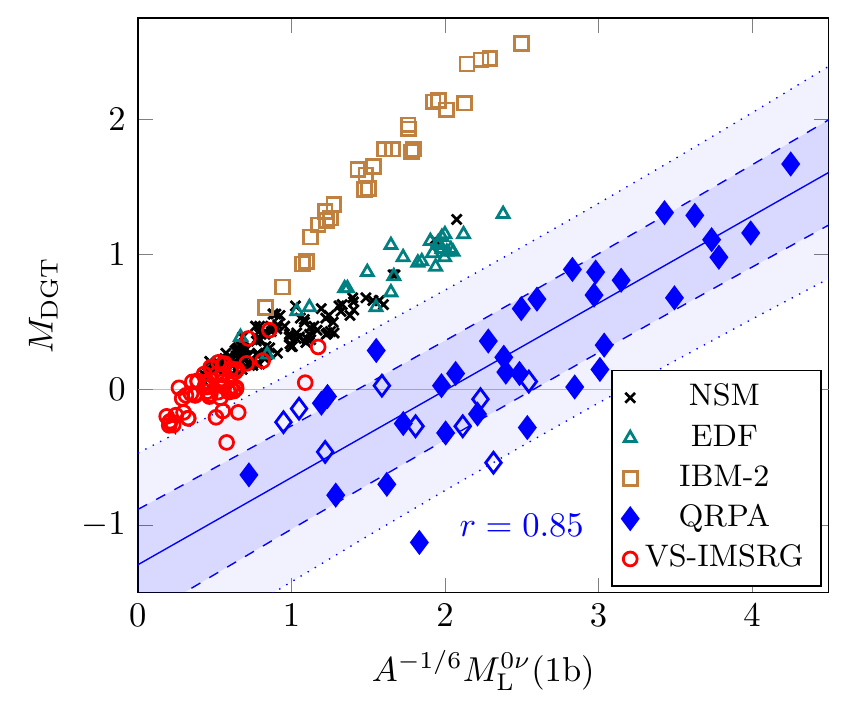}
    \caption{$0\nu\beta\beta$-decay NME (scaled by $A^{-1/6}$) vs double Gamow-Teller NME, obtained with different many-body methods (see text). The QRPA results include NMEs obtained in the present work (solid diamonds) and those of Ref.~\cite{Simkovic2011} (open diamonds). 
    Only the former are used for the linear fit (solid blue line) and 68\% and 95\% CL prediction bands (dashed and dotted blue lines, respectively).}
    \label{fig:M(0vbb)-R-DGT}
\end{figure}

First, we study the relation between $0\nu\beta\beta$ decay and DGT transitions. Figure \ref{fig:M(0vbb)-R-DGT} shows DGT NMEs against $0\nu\beta\beta$-decay NMEs. To be consistent with earlier calculations performed in the NSM \cite{Shimizu2018} (black crosses), EDF \cite{Rodriguez2013} (green triangles), IBM-2 \cite{Barea2015} (brown squares), and VS-IMSRG \cite{Yao22,Belley2022} (red circles) we do not include the contributions from 2BCs or the short-range operator to the $0\nu\beta\beta$-decay NMEs. All results cover only isospin-changing transitions. We compensate the mass-dependence of the $0\nu\beta\beta$-decay NMEs, coming on the one hand from $R$ in Eq. \eqref{eq:H_K} and on the other hand from the two-body operators evaluated in the harmonic oscillator basis, with a scaling factor $A^{-1/6}$ as in Ref.~\cite{Brase2022}. Note that in the VS-IMSRG, however a better correlation is observed when scaling the $0\nu\beta\beta$-decay NMEs by a factor $A^{-1/3}$~\cite{Yao22,Belley2022}, because this method does not depend on the harmonic-oscillator basis. The results obtained in the present work (solid blue diamonds) cover pnQRPA calculations with $0.6\leq g_{\rm pp}^{T=0}\leq 0.8$ as well as results obtained by adjusting $g_{\rm pp}^{T=0}$ to measured $2\nu\beta\beta$-decay half-lives. Figure \ref{fig:M(0vbb)-R-DGT} also includes the pnQRPA NMEs calculated in Ref.~\cite{Simkovic2011} (open blue diamonds) with $g_{\rm pp}$'s adjusted to observed $2\nu\beta\beta$ decays (these are not used in the fit), which correspond to the 'QRPA' NMEs shown in Fig.~4 of Ref.~\cite{Shimizu2018} --- however here we do not take their absolute value. These results differ from our NMEs obtained with $g_{\rm pp}$ adjusted to $2\nu\beta\beta$ decay because of the different two-body interaction, single-particle basis and values of the $g_{\rm ph}$, $g_{\rm pp}^{T=1}$ parameters used.

Figure \ref{fig:M(0vbb)-R-DGT} shows a good linear correlation between the pnQRPA $0\nu\beta\beta$-decay and DGT NMEs (solid blue line)
\begin{equation}
    M_{\rm DGT}=-1.295 + 0.645A^{-1/6}M^{0\nu}_{\rm L}({\rm 1b})\;,
\end{equation}
with 68\% and 95\% confidence-level (CL) prediction bands shown in Fig.~\ref{fig:M(0vbb)-R-DGT} as the shaded regions between dashed and dotted lines, respectively. The best linear fit has a slope similar to that observed in the other approaches, but is shifted to the right compared to them. Also, the pnQRPA results are more spread than in other many-body methods --- the correlation coefficient is $r=0.85$.  While Fig.~\ref{fig:M(0vbb)-R-DGT} corresponds to $g_{\rm A}=1.27$, similar correlations for effective NMEs $M'^{0\nu}=(g_{\rm A}^{\rm eff}/g_{\rm A})^2M^{0\nu}(g_{\rm A}^{\rm eff})$ can be obtained approximately by multiplying $M^{0\nu}$ by $(g_{\rm A}^{\rm eff}/g_{\rm A})^2$.

\begin{figure}[t]
    \centering
    \includegraphics[width=\linewidth]{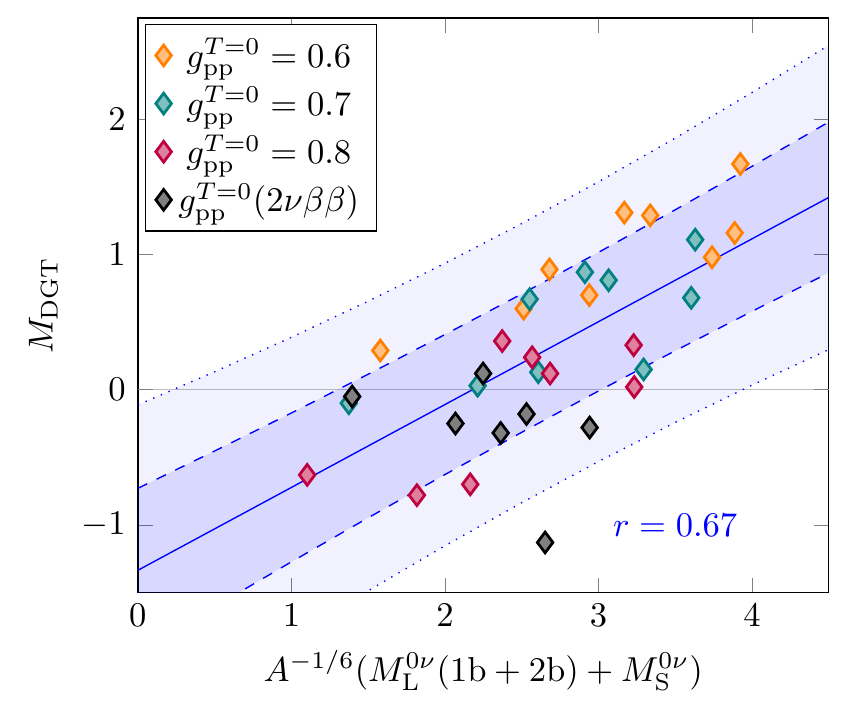}
    \caption{$0\nu\beta\beta$-decay NME (scaled by $A^{-1/6}$) vs double Gamow-Teller NME, obtained with different $g_{\rm pp}^{T=0}$ values. $0\nu\beta\beta$-decay NMEs include two-body currents and the short-range NME. Black diamonds correspond to $g_{\rm pp}^{T=0}$ values adjusted to measured $2\nu\beta\beta$ decays with $g_{\rm A}=1.27$. Blue lines show the best linear fit (solid) and 68\% and 95\% CL prediction bands (dashed and dotted).}
    \label{fig:0vbb-GTcl}
\end{figure}

The difference between the pnQRPA correlation and the one common to other many-body methods can be partly explained by the different role of the $1^+$ multipoles in the $0\nu\beta\beta$-decay NME. For instance, in the NSM $1^+$ states account for $\sim 10-30\%$ of the total NME \cite{Sen'kov2013,Sen'kov2014,Sen'kov2014(R),Neacsu2015}. In contrast, in the pnQRPA the $1^+$ contribution is normally below $10\%$ of the total NME, and depending on the value of $g_{\rm pp}$ the relative sign of the $1^+$ contribution and the total NME can be different~\cite{Simkovic2008,Hyvarinen2015}. Ignoring other multipoles, and comparing just the contribution of $1^+$ states to the NMEs, we observe a very clear correlation (see Fig. \ref{fig:M(0vbb-1+)-R-DGT} in Appendix \ref{sec:1+correlation}).
Also, the correlation observed in the present study, based on the spherical pnQRPA, may change if nuclear deformation is taken into account in the deformed QRPA, which predicts generally smaller NMEs for $0\nu\beta\beta$ decay \cite{Fang2011,Fang2018}. Thus, the correlation could potentially move towards the one observed in other models.

Figure~\ref{fig:0vbb-GTcl} shows the relation between $0\nu\beta\beta$-decay and DGT NMEs when we add 2BCs and the short-range term into the $0\nu\beta\beta$-decay NMEs. The results correspond to the central values of the NMEs considering the range coming from the uncertainty of the 2BCs and the coupling of the contact term \cite{Jokiniemi2022}. To study the correlation more closely, Fig.~\ref{fig:0vbb-GTcl} separates the results obtained with different values of $g_{\rm pp}^{T=0}$. For $0.6\leq g_{\rm pp}^{T=0}\leq0.8$, we observe a linear correlation
\begin{equation}
    M_{\rm DGT}=-1.336 + 0.613A^{-1/6}(M^{0\nu}_{\rm L}({\rm 1b+2b})+M^{0\nu}_{\rm S})\;,
\end{equation}
very similar to the one obtained without 2BCs and the short-range NME, because these two effects largely cancel each other \cite{Jokiniemi2022}. Nonetheless  the correlation coefficient of the fit is worsened to $r=0.67$. This reduction is more marked than in the corresponding correlation between $0\nu\beta\beta$- and $2\nu\beta\beta$-decays NMEs, which changes from $r=0.84$ to $r=0.80$~\cite{Jokiniemi2022}. 

\begin{figure}[t]
\centering
\includegraphics[width=\linewidth]{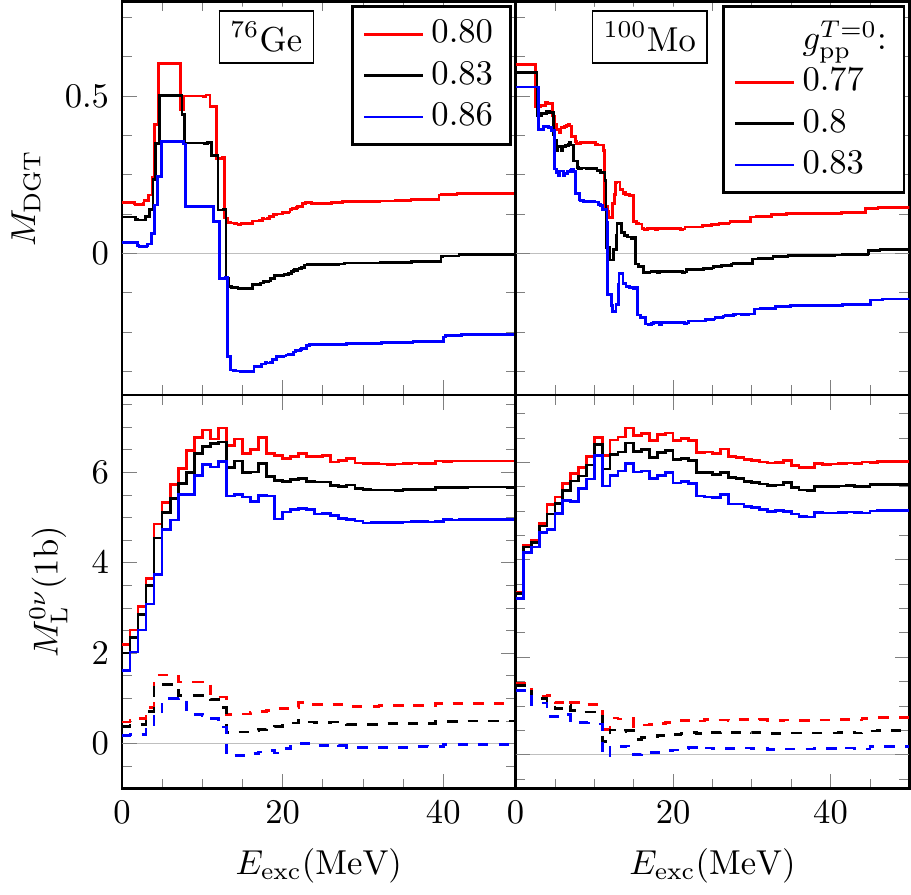}
    \caption{Double Gamow-Teller (upper panel) and $0\nu\beta\beta$-decay (lower panel) NMEs as function of the excitation energy of the intermediate states. In the lower panels, the dashed lines show the contribution from $1^+$ states. 
    }
    \label{fig:DGT-0vbb(E)}
\end{figure}

The relatively small correlation coefficients are related to the small DGT NMEs, especially for $g_{\rm pp}^{T=0}$ fixed to measured $2\nu\beta\beta$ data.
Figure~\ref{fig:0vbb-GTcl} shows that these DGT NMEs are typically close to zero or negative, preventing any visible correlation between NMEs.
The upper panel of Fig. \ref{fig:DGT-0vbb(E)} investigates this further by showing the running sum of the DGT NME for $^{76}$Ge and $^{100}$Mo. These very small NMEs (black lines) result from a strong cancellation in the contributions of low- and high-energy intermediate states. The cancellation is much milder for the red and blue lines, which show the running sums obtained with $g_{\rm pp}^{T=0}$ values 0.03 units below and above, chosen to stay below the pnQRPA breakdown. In these two cases, $M_{\rm DGT}$ varies by $\sim 0.2$ in both nuclei. Even though the energy distributions in these two nuclei look quite different, there are strong negative cancellations at around $10-15$ MeV in both cases, a phenomenon observed in all studied nuclei and also noticed previously \cite{Simkovic2011,Simkovic2018}. These negative contributions in the case of $^{76}$Ge consist mainly of transitions $2{\rm n}0f_{7/2}\rightarrow 2{\rm p}0f_{7/2}$ and ${\rm n}0g_{9/2}{\rm n}0g_{7/2}\rightarrow {\rm p}0g_{9/2}{\rm p}0g_{7/2}$.
There are especially large negative contributions for $^{100}$Mo calculated with $g_{\rm pp}^{T=0}=0.89$ adjusted to $t_{1/2}^{2\nu}$, see the trend in Fig. \ref{fig:DGT-0vbb(E)} for increasing $g_{\rm pp}^{T=0}$ values. In fact, the corresponding NMEs lie below the prediction bands in Figs. \ref{fig:M(0vbb)-R-DGT} and \ref{fig:0vbb-GTcl}, worsening the correlation coefficient.

\begin{figure}[t]
    \centering
    \includegraphics[width=\linewidth]{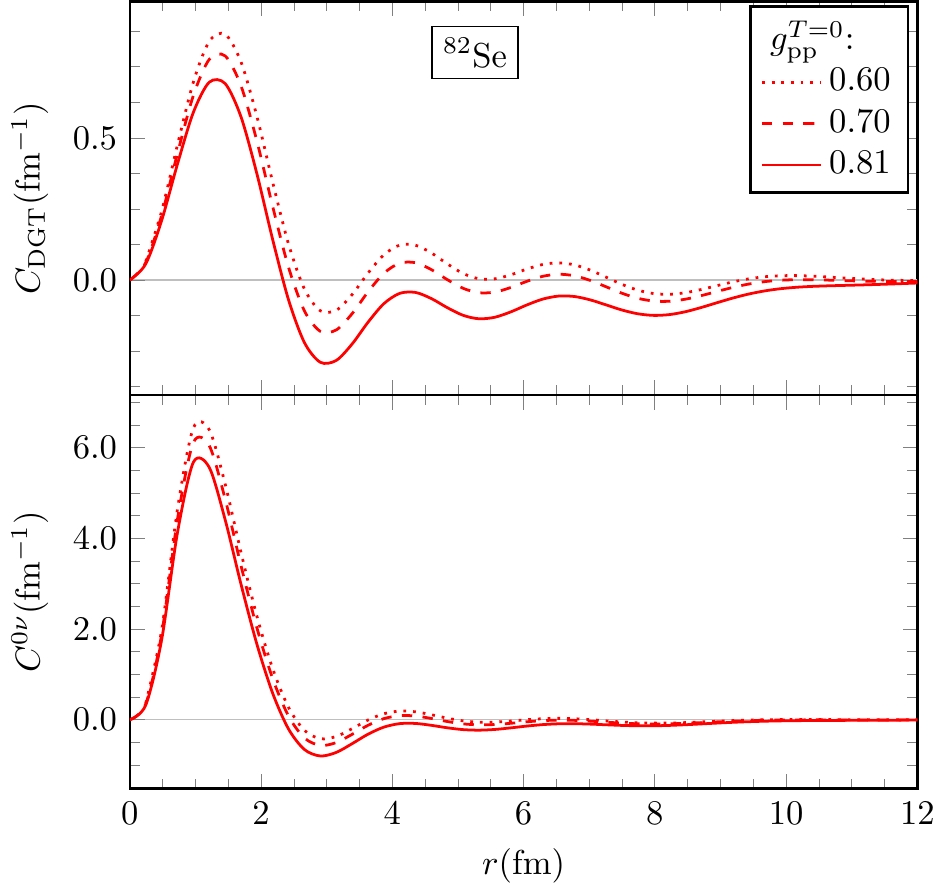}
    \caption{Radial distributions of $M^{0\nu}_{\rm L}({\rm 1b})$ (upper panel) and $M_{\rm DGT}$ (lower panel) in $^{82}$Se for different $g_{\rm pp}^{T=0}$ values.}
    \label{fig:rdep-0vbb-DGT}
\end{figure}

Figure~\ref{fig:DGT-0vbb(E)} also highlights that decreasing the value of $g_{\rm pp}^{T=0}$ mitigates the cancellations observed at the $10-15$-MeV region and slightly increases the low-energy contribution. This common feature for all studied nuclei improves the correlations. The lower panel of Fig. \ref{fig:DGT-0vbb(E)} shows the running sums for $M^{0\nu}$ obtained with the same values of $g_{\rm pp}^{T=0}$. Dashed lines indicate the contribution coming from transitions through $1^+$ intermediate states and solid lines the total NME. $M^{0\nu}$ is much less dependent on the value of $g_{\rm pp}^{T=0}$ than $M_{\rm DGT}$, while the behavior of the $1^+$ contribution, constituting a minor fraction of the total NME, behaves quite like $M_{\rm DGT}$ in the upper panel.

The origin of the correlation between $M^{0\nu}$ and $M_{\rm DGT}$ was first attributed to the relative dominance of short-distance physics in both processes~\cite{Shimizu2018}, which would lead to a correlation between the two NMEs since they share the same spin-isospin structure~\cite{Anderson2010,Bogner2012}. We investigate this by calculating the radial NME distributions $C^{0\nu}(r)$ and $C_{\rm DGT}(r)$ which satisfy
\begin{align}
M^{0\nu}_{\rm L}({\rm 1b})&=\int_0^{\infty} C^{0\nu}(r){\rm d}r\;, \\
M_{\rm DGT}&=\int_0^{\infty}C_{\rm DGT}(r){\rm d}r\;,
\end{align}
and are defined as
\begin{align}
\label{eq:0nu_rad_dens}
&C^{0\nu}(r)=C^{0\nu}_{\rm GT}(r)-C^{0\nu}_{\rm F}(r)+C^{0\nu}_{\rm T}(r)\;, \\
    &C^{0\nu}_{\rm K}(r)= \nonumber \\ &\sum_{k,ab}(0^+_f||\mathcal{O}^K_{ab}\tau^-_a\tau^-_bH_K(r_{ab})f_{\rm SRC}^2(r_{ab})\delta(r-r_{ab})||0^+_i)\;,
\end{align}
and 
\begin{equation}
 \label{eq:dgt_rad_dens}
    C_{\rm DGT}(r)=\frac{1}{\sqrt{3}}\sum_{k,ab}(0^+_f||\boldsymbol{\sigma}_a\cdot\boldsymbol{\sigma}_b\tau^-_a\tau^-_b\delta(r-r_{ab})||0^+_i)\;. 
\end{equation}

Figure~\ref{fig:rdep-0vbb-DGT} shows the radial distributions of the $0\nu\beta\beta$-decay (without the short-range NME or 2BCs) and DGT NMEs in $^{82}$Se. The results correspond to $g_{\rm pp}^{T=1}$ adjusted so that $M^{2\nu}_{\rm F}=0$  (see Table \ref{tab:gpps-2vbb}) and different $g_{\rm pp}^{T=0}$ values. For $g_{\rm pp}^{T=0}=0.81$, we have $M_{\rm DGT}\approx0$: the positive short-range contribution gets cancelled by the negative long-distance tail. These cancellations due to long-range contributions have also been shown to deteriorate the linear correlation between NMEs in {\it ab initio} calculations~\cite{Yao22}. For smaller $g_{\rm pp}^{T=0}$ values, however, Fig.~\ref{fig:rdep-0vbb-DGT} shows that the short-range bump slightly increases, while the negative tail gets notably less prominent, so that short-range contributions are dominant as in the $0\nu\beta\beta$-decay distributions computed with the same values of $g_{\rm pp}^{T=0}$. A linear correlation can then be expected in these cases. The $C^{0\nu}$ distribution is much less dependent on $g_{\rm pp}^{T=0}$ due to the minor role of the $1^+$ multipole.
We note that even though the DGT NME involves only intermediate $1^+$ states, the radial distribution gets contributions from all intermediate multipoles due to $\delta(r-r_{ab})$ in the definition of $C_{\rm DGT}$. However, when integrated over $r$, contributions coming from multipoles other than $1^+$ vanish~\cite{Simkovic2011}.

\section{Correlation Between $\gamma\gamma$ and $0\nu\beta\beta$ Decay NMEs}

\begin{figure}[t!]
    \centering
    \includegraphics{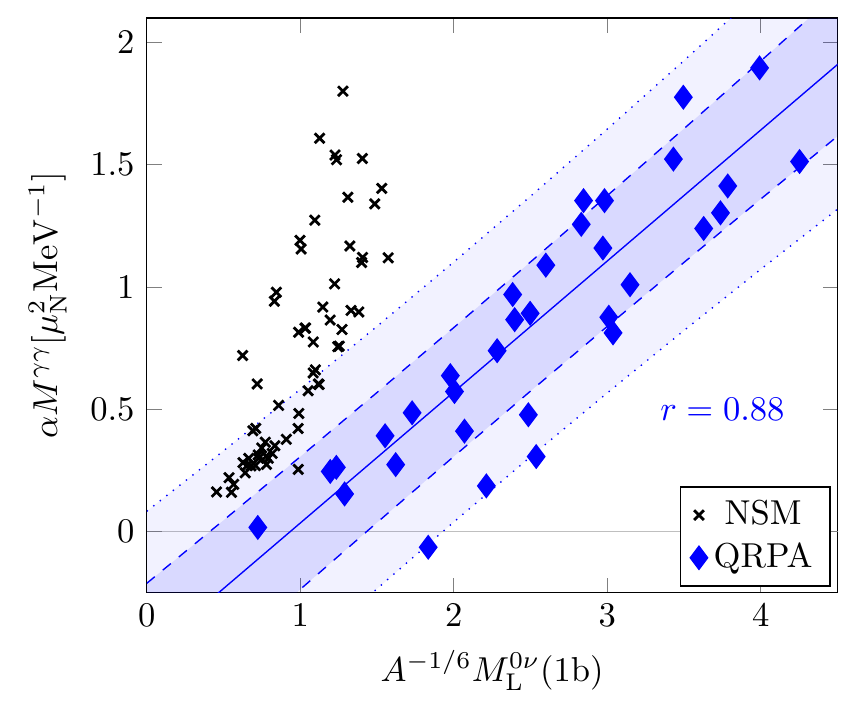}
    \caption{$0\nu\beta\beta$-decay NME (scaled by $A^{-1/6}$) vs $\gamma\gamma$-M1M1 NME, comparing present QRPA results to NSM ones from Ref.~\cite{Romeo2021}. Blue lines show the best linear fit (solid) and 68\% and 95\% CL prediction bands (dashed and dotted) for the QRPA results.}
    \label{fig:M(0vbb)-M(M1M1)-QRPA-NSM}
\end{figure}

\begin{figure}[t!]
    \centering
    \includegraphics[width=\linewidth]{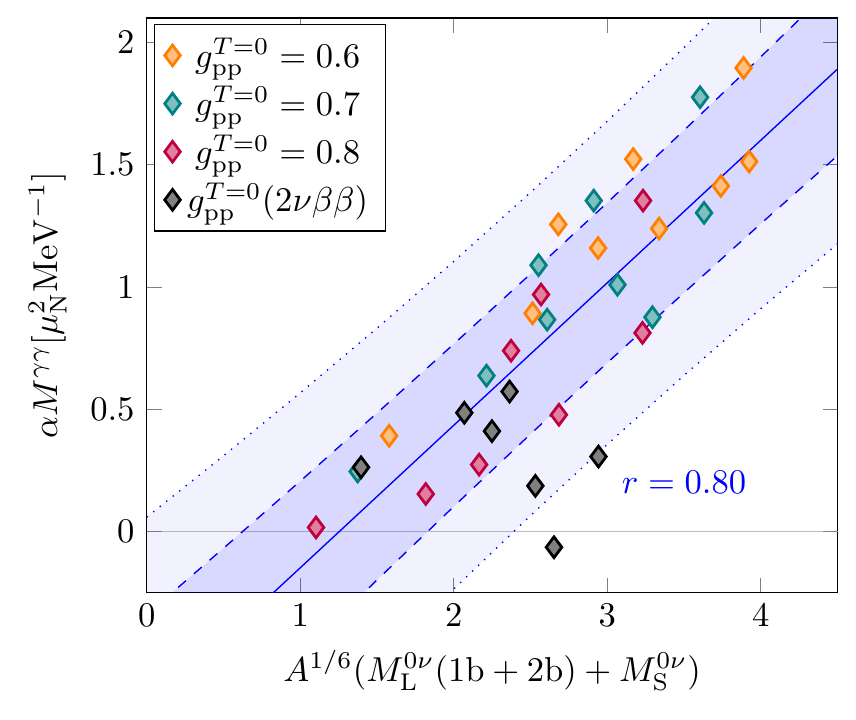}
    \caption{$0\nu\beta\beta$-decay NME (scaled by $A^{-1/6}$) vs $\gamma\gamma$-M1M1 NME, obtained with different $g_{\rm pp}^{T=0}$ values. $0\nu\beta\beta$-decay NMEs include two-body currents and the short-range term. Black diamonds correspond to $g_{\rm pp}^{T=0}$ values adjusted to measured $2\nu\beta\beta$ decays with $g_{\rm A}=1.27$. Blue lines show the best linear fit (solid) and 68\% and 95\% CL prediction bands (dashed and dotted).} 
    \label{fig:M(0vbb)-M(M1M1)}
\end{figure}

Figure \ref{fig:M(0vbb)-M(M1M1)-QRPA-NSM} shows the relation between $\gamma\gamma$-M1M1 and long-range $0\nu\beta\beta$-decay NMEs in the same manner as in Fig.~\ref{fig:M(0vbb)-R-DGT}. Again, we observe a good linear correlation when including results for different values of $g_{\rm pp}^{T=0}$. The best fit is given by
\begin{equation}
A^{1/6}\alpha M^{\gamma\gamma}=-1.265+0.557M^{0\nu}_{\rm L}({\rm 1b})\;,
    \label{eq:0vbb-M1M1-corr}
\end{equation}
with correlation factor $r=0.88$. Hence, in the pnQRPA framework the correlation between $\gamma\gamma$  and $0\nu\beta\beta$-decay NMEs is better than the one between DGT and $0\nu\beta\beta$-decay NMEs, contrary to the NSM (see Refs. \cite{Shimizu2018} and \cite{Romeo2021}). This is because in the case of M1M1 transitions the energy denominator in Eq.~\eqref{eq:M_M1M1_QRPA} prevents the strong cancellations between low- and high-energy intermediate-state contributions that can occur in DGT transitions. In fact, the correlation is also slightly stronger than the one observed between pnQRPA $0\nu\beta\beta$- and $2\nu\beta\beta$-decay NMEs~\cite{Jokiniemi2022}.

\begin{figure}[t!]
    \centering
    \includegraphics[width=\linewidth]{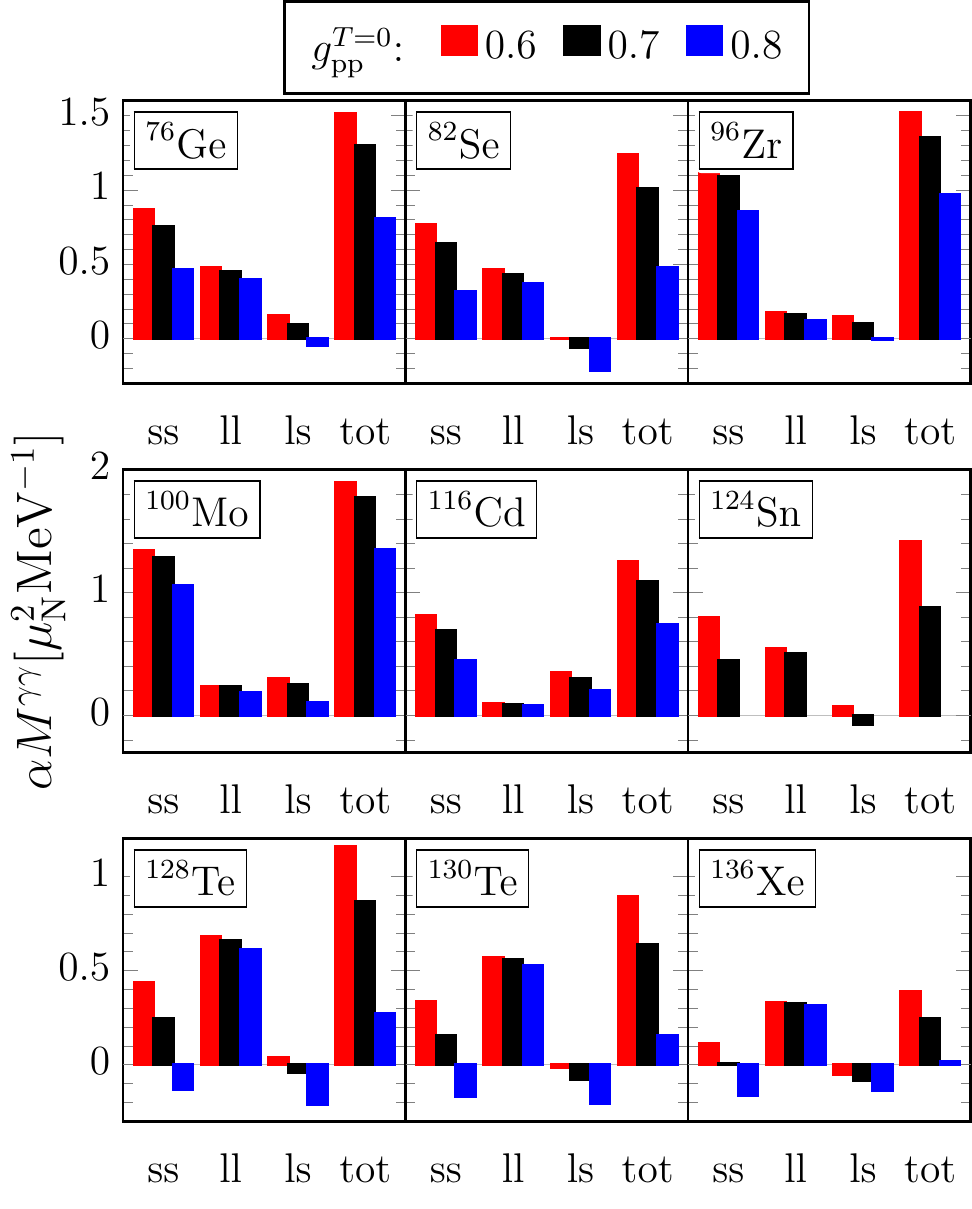}
    \caption{$\gamma\gamma$-M1M1 NME (tot) decomposed into spin (ss), orbital (ll) and interference (ls) terms, for different $g_{\rm pp}^{T=0}$ values.}
    \label{fig:M1M1-decomposition}
\end{figure}

Figure~\ref{fig:M(0vbb)-M(M1M1)-QRPA-NSM} also compares the correlation found in the present work  with the one found in the NSM~\cite{Romeo2021}. While both many-body methods indicate a correlation between the NMEs, each framework finds a different one: the slope of the linear fit in the pnQRPA is about half the one observed in the NSM. Also, the pnQRPA best fit is shifted to the right, however not as prominently as in Fig.~\ref{fig:M(0vbb)-R-DGT}. Again, this could be due to the different role of $1^+$ and other multipoles in both methods and might change if deformation would be included into the pnQRPA calculations. If only the $1^+$ contribution to the pnQRPA $0\nu\beta\beta$-decay NMEs is included, the situation resembles that of the DGT transitions and $1^+$ $0\nu\beta\beta$-decay multipoles, the pnQRPA correlation lying on the left side from the NSM one  (see Fig. \ref{fig:M(0vbb-1+)-M(M1M1)} in Appendix \ref{sec:1+correlation}). However, the correlation is much weaker in this case due to the orbital angular momentum operator and the energy denominator involved in the $\gamma\gamma$-M1M1 NMEs.

Figure~\ref{fig:M(0vbb)-M(M1M1)} shows the correlation after adding 2BCs and the short-range term into the $0\nu\beta\beta$-decay NMEs. We find the best fit
\begin{equation}
 A^{1/6}\alpha M^{\gamma\gamma}=-0.732+0.583(M^{0\nu}_{\rm L}({\rm 1b+2b})+M^{0\nu}_{\rm S})\,,
    \label{eq:0vbb-M1M1-corr-2bcs-contact}   
\end{equation}
with correlation coefficient $r=0.80$, which is notably stronger than the corresponding one between $0\nu\beta\beta$ and DGT NMEs. This is likely so because of the absence of cancellations in the $\gamma\gamma$-M1M1 NME running sums due to the energy denominator.

\begin{figure}[t]
\centering
\includegraphics[width=\linewidth]{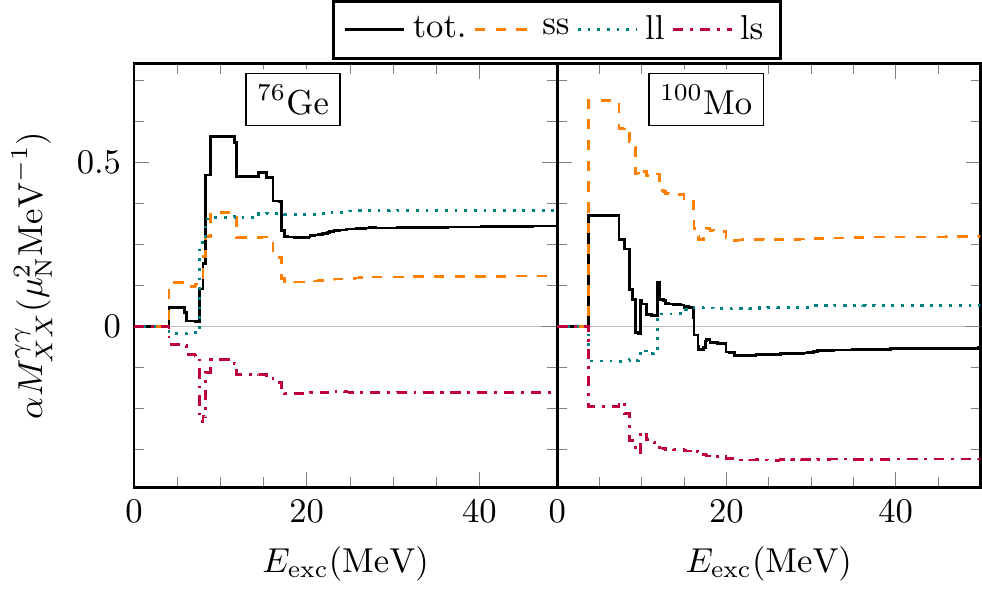}
    \caption{Spin (ss), orbital (ll) and interference (ls) parts of $\gamma\gamma$-M1M1 NMEs as a function of the excitation energy of the intermediate states, with $g_{\rm pp}^{T=0}$ adjusted to $2\nu\beta\beta$-decay data.}
    \label{fig:M1M1-parts(E)}
\end{figure}

In order to further study the $\gamma\gamma$-M1M1 NMEs, we decompose them into spin, orbital and interference parts as \cite{Romeo2021}
\begin{equation}
M^{\gamma\gamma}=M^{\gamma\gamma}_{\rm ss}+M^{\gamma\gamma}_{\rm ll}+M^{\gamma\gamma}_{\rm ls}\;.
\label{eq:M1_lsdec}
\end{equation}
Figure~\ref{fig:M1M1-decomposition} shows the three contributions for all studied nuclei for $g_{\rm pp}^{T=0}=0.6$, $g_{\rm pp}^{T=0}=0.7$ and $g_{\rm pp}^{T=0}=0.8$. The spin part dominates in the lighter nuclei with $A<124$, but the orbital part becomes dominant in heavier nuclei with $A\geq 124$. Furthermore, Fig.~\ref{fig:M1M1-decomposition} highlights that the orbital part is indeed much less sensitive to the value of $g_{\rm pp}^{T=0}$ than the parts containing the spin operator. Except in heavier nuclei with large values of $g_{\rm pp}^{T=0}$, the spin and orbital parts carry the same sign. This explains why the correlation is present even in cases where the orbital part is dominant. The interference term, in turn, often carries the opposite sign of the leading contributions to the total NME. If this part would be excluded, the correlation between M1M1 and $0\nu\beta\beta$ NMEs would improve.

Figure~\ref{fig:M1M1-parts(E)} shows the different parts of the $\gamma\gamma$-M1M1 NMEs in Eq.~\eqref{eq:M1_lsdec} for $^{76}$Ge and $^{100}$Mo as a function of the excitation energy of the intermediate states of the transition, for $g_{\rm pp}^{T=0}$ values adjusted to measured $2\nu\beta\beta$ data. Figure~\ref{fig:M1M1-parts(E)} illustrates that, for these nuclei, the spin part drives the overall behavior of the total NME, but the orbital and interference terms are important for the final NME value. The orbital part gets contribution mainly from a few $1^+$ states at $E_{\text{exc}}\sim10$ MeV for $^{76}$Ge and $E_{\text{exc}}\sim15$  MeV for $^{100}$Mo. In the case of $^{76}$Ge, these contributions consist mainly of transitions $2{\rm n}0f_{5/2}\rightarrow 2{\rm p}0f_{5/2}$ and $2{\rm n}0g_{9/2}\rightarrow 2{\rm p}0g_{9/2}$, and in the case of $^{100}$Mo of transitions $2{\rm n}0h_{11/2}\rightarrow 2{\rm p}0h_{11/2}$ and ${\rm n}0g_{9/2}{\rm n}0g_{7/2}\rightarrow {\rm p}0g_{9/2}{\rm p}0g_{7/2}$. The spin term is more evenly distributed at excitation energies $E_{\text{exc}}\approx 5-15\,{\rm MeV}$. This behavior of the pnQRPA $\gamma\gamma$-M1M1 NMEs is somewhat different to the running sum in NSM calculations, which for these nuclear masses are typically dominated by one or few intermediate states at $E_{\text{exc}}\lesssim 10\,{\rm MeV}$~\cite{Romeo2021}. The higher energies relevant for the $\gamma\gamma$-M1M1 NMEs explain the smaller slope of the pnQRPA correlation with respect to the NSM one in Fig.~\ref{fig:M(0vbb)-M(M1M1)-QRPA-NSM}, driven by the energy denominator in Eq.~\eqref{eq:M_M1M1}.

\begin{figure}[t]
    \centering
    \includegraphics[width=\linewidth]{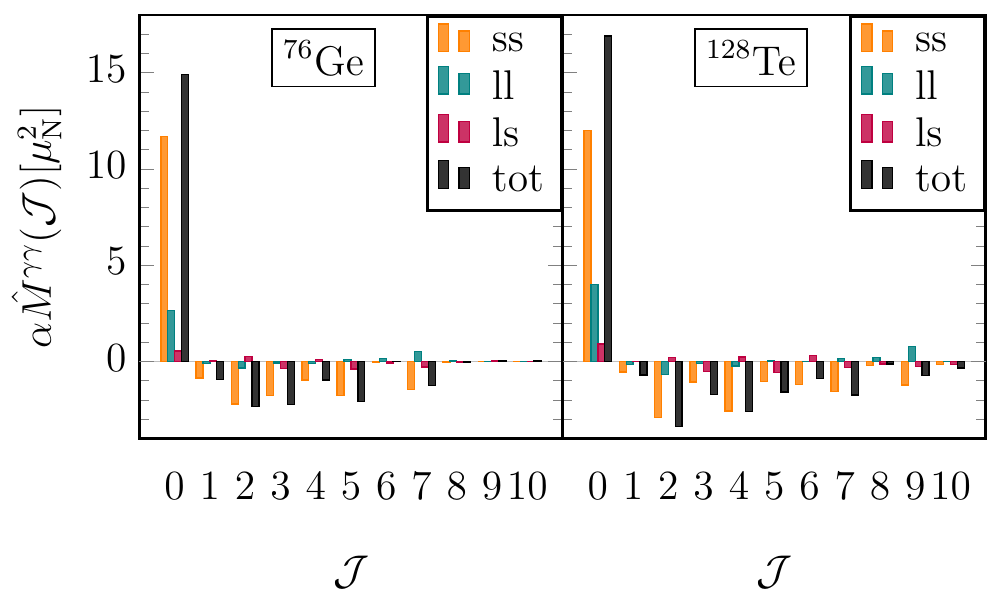}
    \caption{Decomposition of the numerator of the $\gamma\gamma$-M1M1 NMEs in terms of the two-nucleon total angular momenta $\mathcal{J}$, obtained for $g_{\rm pp}^{T=0}=0.7$.}
    \label{fig:M1M1-J-pair}
\end{figure}

In order to better understand the origin of the correlation between $\gamma\gamma$ and $0\nu\beta\beta$ NMEs, we decompose the numerator $\Hat{M}^{\gamma\gamma}$ of the $\gamma\gamma$-M1M1 NMEs --- without the energy denominator in Eq.~\eqref{eq:one-body_NME} ---  in terms of the two-body nucleon total angular momenta $\mathcal{J}$ of the decaying nucleons by utilising Eq.~\eqref{eq:pnQRPA-NMEs} with the $\mathbf{M1\cdot M1}$ operator. It is worth noting that while M1M1 transitions only run through $1^+$ intermediate states, all $J^{\pi}$ multipoles contribute to the $\mathcal{J}$ decomposition --- however, $J^{\pi}\neq 1^+$ contributions vanish when summed over $\mathcal{J}$. Figure~\ref{fig:M1M1-J-pair} shows the decomposition for  $^{76}$Ge and $^{128}$Te. The qualitative behavior is similar to the corresponding NSM decomposition in Fig. 4 of Ref.~\cite{Romeo2021}: the $\gamma\gamma$-M1M1 NME is clearly driven by $\mathcal{J}=0$ pairs, while the higher-momenta pairs reduce the total value of the NME. Thus the decomposition resembles that of the $0\nu\beta\beta$ NMEs \cite{Simkovic2008,Menendez2009}. Furthermore, the spin part is clearly dominating the leading $\mathcal{J}=0$ contribution to $\hat{M}^{\gamma\gamma}$. Nonetheless, since the orbital part only gets mildly cancelled by the $\mathcal{J}\neq 0$ pairs, the total values of the spin and orbital parts can be comparable, see Fig.~\ref{fig:M1M1-decomposition}. 

\begin{figure}[t!]
    \centering
    \includegraphics[width=\linewidth]{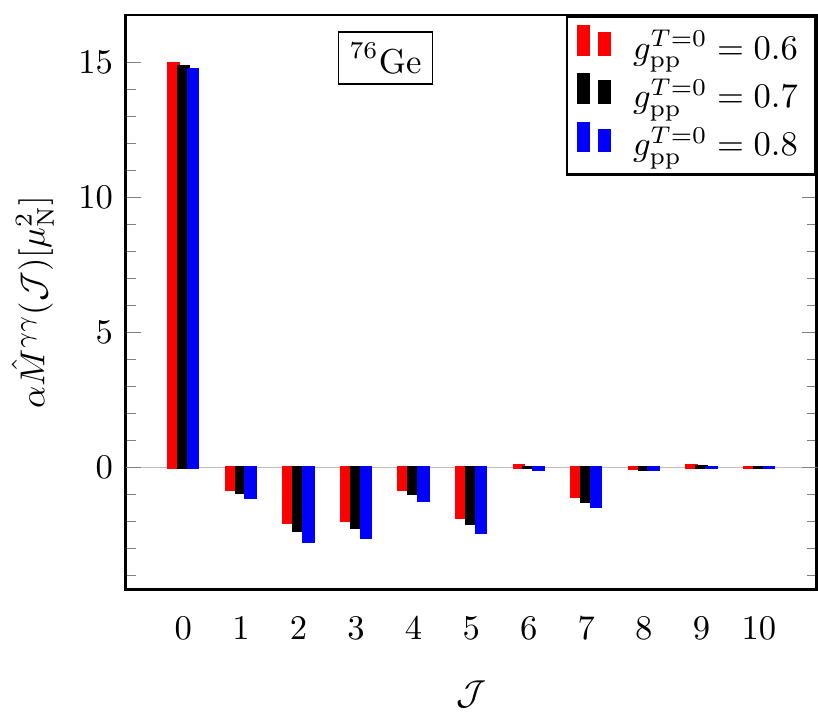}
    \caption{Decomposition of the numerator of the $\gamma\gamma$-M1M1 NME in terms of the two-nucleon angular momenta $\mathcal{J}$ for calculations using different $g_{\rm pp}^{T=0}$ values.}
    \label{fig:M1M1-J-pair-gpp}
\end{figure}

Finally, to gain deeper understanding on the $g_{\rm pp}$-dependence of the $\gamma\gamma$-M1M1 NME, Fig. \ref{fig:M1M1-J-pair-gpp} shows the $\mathcal{J}$ decomposition of the numerator of the $\gamma\gamma$-M1M1 NME in $^{76}$Ge obtained with different values of $g_{\rm pp}^{T=0}$. Figure \ref{fig:M1M1-J-pair-gpp} indicates that increasing the value of $g_{\rm pp}^{T=0}$ increments the (negative) contribution of $\mathcal{J}\neq 0$ pairs hence decreasing the total value of the NME. Figure \ref{fig:M1M1-decomposition} shows that the $g_{\rm pp}^{T=0}$ dependence of $\hat{M}^{\gamma\gamma}$ is mostly coming from the spin part.

\section{Conclusions}

Linear correlations between $0\nu\beta\beta$ decays and standard-model-allowed two-nucleon transfer reactions have previously been found in the literature: between $0\nu\beta\beta$ and DGT transitions in the nuclear shell model, energy-density-functional theory \cite{Shimizu2018}, the interacting boson model~\cite{Santopinto2018,Barea2015} and in different \emph{ab initio} frameworks \cite{Yao22,Weiss:2021rig,Belley2022}; and also between $0\nu\beta\beta$ and M1M1 in the nuclear shell model \cite{Romeo2021}. On the contrary, no correlation has been found in the pnQRPA approach \cite{Simkovic2018,Lv2023}.

In the present study, we perform systematic calculations on DGT and M1M1 transitions and $0\nu\beta\beta$ decays to further study the relations between these processes in the pnQRPA. When exploring a wide range of proton-neutron pairing strength values, covering the typical range of values that describe well $\beta\beta$- and $\beta$-decay data, we find good linear correlations between $0\nu\beta\beta$ decays and both DGT and M1M1 transitions. Our findings are in contrast with previous pnQRPA studies, performed with fixed proton-neutron pairing strengths, which did not find any apparent correlation between $0\nu\beta\beta$-decay and DGT NMEs. The discrepancy with previous studies can be explained by cancellations in the running sums of the DGT NMEs occurring in the vicinity of the fixed pairing strengths studied in these works. In fact, the results of some previous pnQRPA calculations \cite{Simkovic2011} fall inside our prediction bands obtained by exploring a wide range of proton-neutron pairing strengths.
We also note that in Ref. \cite{Lv2023}, no correlation was found by varying the particle-hole and isovector particle-particle pairing strengths for spherical QRPA calculations based on Skyrme functionals. However, that work does not explore variations in the isoscalar particle-particle pairing channel in this context.

The pnQRPA correlations found in this work differ from the correlations found using different nuclear-theory frameworks, even though in the correlation with DGT transitions the slopes of the best-fit functions are comparable. This difference might be related to the different contributions of intermediate states to $0\nu\beta\beta$-decay NMEs in the pnQRPA or to the lack of deformation in our pnQRPA calculations.  We also find good linear correlations  when including two-body currents and the short-range $0\nu\beta\beta$-decay NME. However, when doing so the correlation between DGT transitions and $0\nu\beta\beta$ decays is weakened. For M1M1 transitions we find that NMEs dominated by the orbital angular momentum part of the operator show also a good correlation with $0\nu\beta\beta$ decays NMEs. Compared to the NSM, the pnQRPA correlation has a smaller slope because the intermediate states contributing to M1M1 transitions have typically higher energies. In sum, our findings suggest that if DGT and M1M1 transitions are measured, their relations with $0\nu\beta\beta$ decay could help to constrain the uncertain values of $0\nu\beta\beta$-decay NMEs.

\begin{acknowledgments}
We thank A. Belley for sharing the VS-IMSRG results and for helpful discussions and B. Romeo for discussions and the careful reading of the manuscript.
This work was supported by the Finnish Cultural Foundation grant No. 00210067, the Arthur B. McDonald Canadian Astroparticle Physics Research Institute, and by the ``Ram\'on y Cajal'' program with grant RYC-2017-22781 and grants CEX2019-000918-M and PID2020-118758GB-I00 funded by MCIN/AEI/10.13039/501100011033 and by ``ESF Investing in your future''.
TRIUMF receives funding via a contribution through the National Research Council of Canada.
\end{acknowledgments}

\bibliography{bibs}

\appendix
\section{Correlations between $M^{0\nu}(1^+)$ and $M_{\rm DGT}$ and $M^{\gamma\gamma}$}
\label{sec:1+correlation}

\begin{figure}[b]
    \centering
    \includegraphics[width=\linewidth]{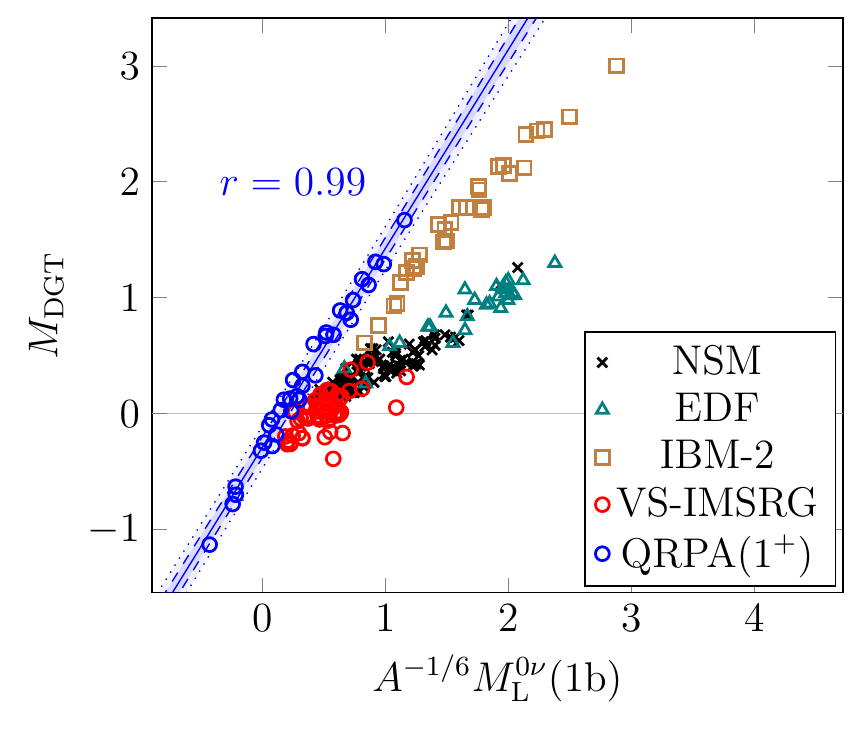}
    \caption{Same as Fig. \ref{fig:M(0vbb)-R-DGT}, but the QRPA results include only the contribution $M^{0\nu}(1^+)$ instead of the total $M^{0\nu}$.}
    \label{fig:M(0vbb-1+)-R-DGT}
\end{figure}

\begin{figure}[b]
    \centering
    \includegraphics[width=\linewidth]{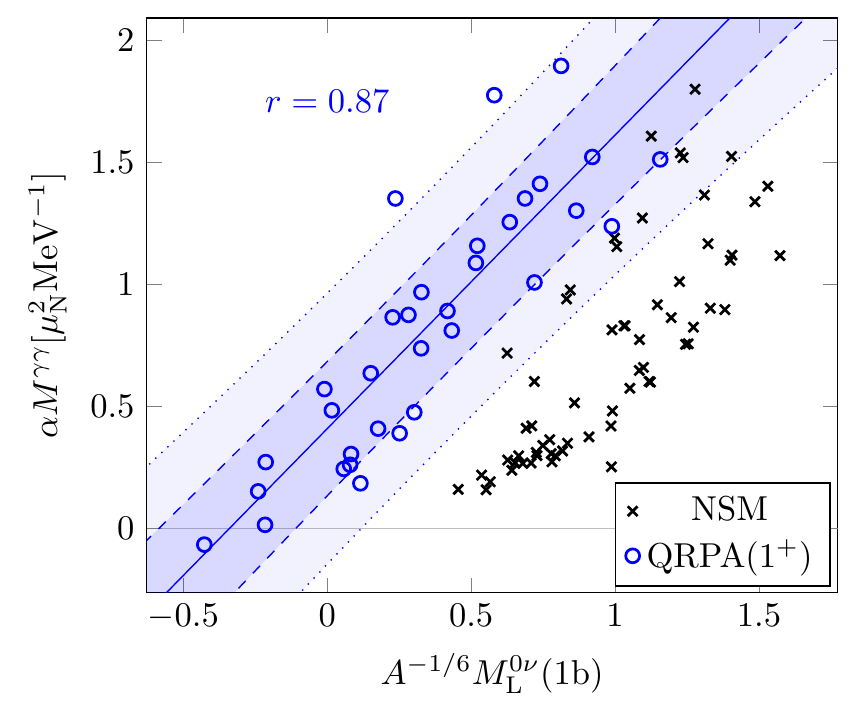}
    \caption{Same as Fig. \ref{fig:M(0vbb)-M(M1M1)-QRPA-NSM} but the QRPA results consider only the contribution $M^{0\nu}(1^+)$ instead of the total $M^{0\nu}$.}
    \label{fig:M(0vbb-1+)-M(M1M1)}
\end{figure}

Figures \ref{fig:M(0vbb-1+)-R-DGT} and \ref{fig:M(0vbb-1+)-M(M1M1)} show the relation between DGT and $\gamma\gamma$-M1M1 NMEs and long-range $0\nu\beta\beta$-decay NMEs but only taking into account the contribution from $1^+$ intermediate states. Figure \ref{fig:M(0vbb-1+)-R-DGT} also shows the correlations between the total $M^{0\nu}_{\rm L}({\rm 1b})$, containing all possible multipole contributions, obtained with other nuclear many-body methods. Figure \ref{fig:M(0vbb-1+)-R-DGT} reveals that the correlation taking into account only the $1^+$ contribution in the QRPA becomes significantly stronger than the one taking into account all other multipoles as well, see Fig. \ref{fig:M(0vbb)-R-DGT}. This is expected because DGT NMEs only receive contributions from $1^+$ intermediate states. Furthermore, the correlation best fit shifts closer to the results obtained with other nuclear models, even though with a steeper slope. In contrast, Fig. \ref{fig:M(0vbb-1+)-M(M1M1)} shows that in the case of M1M1 decays the correlation does not become remarkably better as we take only the $1^+$ contributions into account. This is partly due to the energy denominator in the $\gamma\gamma$-M1M1 NMEs, which introduces some sensitivity on the excitation energies of the $1^+$ states, absent in $0\nu\beta\beta$-decay NMEs because of its very large momentum transfer $p\sim100$ MeV. In addition, the correlation does not improve because $\gamma\gamma$-M1M1 NMEs involve a different operator than DGT or $0\nu\beta\beta$ ones, namely the orbital angular momentum. 

\end{document}